\documentclass[aip,amsmath,preprint]{revtex4-2}

\usepackage{graphicx}
\usepackage{bm}

\usepackage[utf8]{inputenc}
\usepackage[T1]{fontenc}
\usepackage{mathptmx}
\usepackage{etoolbox}
\usepackage{xcolor}

\usepackage{xr}

\makeatletter
\def\@email#1#2{%
 \endgroup
 \patchcmd{\titleblock@produce}
  {\frontmatter@RRAPformat}
  {\frontmatter@RRAPformat{\produce@RRAP{*#1\href{mailto:#2}{#2}}}\frontmatter@RRAPformat}
  {}{}
}%
\makeatother
\begin{document}

\preprint{AIP/123-QED}

\title{Exact analytical solution for the Sakiadis boundary layer}
\author{N. S. Barlow}
\affiliation{School of Mathematics \& Statistics, Rochester Institute of Technology, Rochester, NY, 14623, USA}
\email[corresponding author: ]{nsbsma@rit.edu}
\author{W. C. Reinberger}%
\affiliation{School of Mathematics \& Statistics, Rochester Institute of Technology, Rochester, NY, 14623, USA}
\author{S. J. Weinstein}
\affiliation{School of Mathematics \& Statistics, Rochester Institute of Technology, Rochester, NY, 14623, USA}
\affiliation{Department of Chemical Engineering, Rochester Institute of Technology, 
Rochester, NY, 14623, USA}

\date{\today}

\begin{abstract}
It has recently been established [Naghshineh et al., IMA J. of Appl. Math., \textbf{88}, 1 (2023)] that a convergent series solution may be obtained for the Sakiadis boundary layer problem once key parameters are determined  iteratively using the series itself. Here, we provide exact and explicit analytical expressions for these parameters, including that associated with wall shear, thus completing the exact analytical solution. The complete exact analytical solution to the Sakiadis problem is summarized here for direct use.  
\end{abstract}

\maketitle

The Sakiadis boundary layer along a moving wall is an important flow field in configurations where thin liquid films are coated onto moving substrates~\cite{Weinstein2004}. From the time since~\citet{Sakiadis} applied Blasius's similarity transform to Prandtl's boundary layer equations (with appropriate boundary conditions) to arrive at the similarity solution, the work has been cited nearly 3000 times (>500 times in 2023 alone).  The similarity transformation reduces the nonlinear PDE system (the boundary layer equations) to a third-order nonlinear ODE for which no general analytical solution is known. Hence, it is not surprising that, even in the most recent studies~\cite{Hattori}, the exact solution is obtained numerically. Asymptotically consistent approximant~\cite{Barlow:2017} and convergent series~\cite{FlatWallSakiadisPaper} solutions may be obtained for the Sakiadis problem once key parameters are determined iteratively--using the approximant or series itself\footnote{The same approach was used to solve the analogous non-Newtonian problem~\cite{NN}.}. Here, we provide exact and explicit analytical expressions for these parameters, thus providing---in aggregate---the exact and explicit  analytical solution to the Sakiadis problem itself. 

The Sakiadis initial/boundary value problem to find the similarity solution for the stream function, $f(\eta)$, is given as~\cite{Sakiadis}
\begin{subequations}
\begin{equation}
    2f'''+ff''=0,~0\le\eta<\infty,
    \label{eq:SakODE}
\end{equation}
\begin{equation}
    f(\eta=0)=0,
    \label{eq:SakBC1}
\end{equation}
\begin{equation}
    f'(\eta=0)=1,
    \label{eq:SakBC2}
\end{equation}
\begin{equation}
    f'(\eta\to\infty)=0.
    \label{eq:Sakinf}
\end{equation}
\label{eq:SakBVP}
\end{subequations}
The wall shear parameter, $\kappa$, is extracted from the solution to~(\ref{eq:SakBVP}) as
\begin{equation}
\kappa=f''(\eta=0).
\label{eq:SakBC3}
\end{equation}
This parameter has historically been obtained by either applying a numerical shooting technique~\cite{Cortell} to~(\ref{eq:SakBVP}) or by iterating on a series solution~\cite{Barlow:2017,FlatWallSakiadisPaper}, as cited above. Once $\kappa$ is known, the power series solution about $\eta=0$ may be constructed; however, this series diverges at $\eta\approx4$ due to non-physical singularities in the complex $\eta$-plane~\cite{Barlow:2017,FlatWallSakiadisPaper}.  On the other hand,~\citet{FlatWallSakiadisPaper} show that if the expansion is taken about the other end of the physical domain,  $\eta=\infty$, as is done using the method of dominant balance in~\citet{Barlow:2017}, an infinite series is obtained that converges over the entire physical domain $0\le\eta\le\infty$; it takes the form
\begin{subequations}
\begin{equation}
\frac{f}{C}=1+\gamma e^{-C\eta/2} +\frac{1}{4}\left(\gamma e^{-C\eta/2}\right)^2+\frac{5}{72}\left(\gamma e^{-C\eta/2}\right)^3+\dots,
\label{eq:SakAsymp}
\end{equation}
\begin{equation}
    C>0,
    \label{eq:C}
\end{equation}
\label{eq:dombal}
\end{subequations}
where $C$ and $\gamma$ are constants to be determined.  Note that the constraint~(\ref{eq:C}) is required to be self-consistent with the boundary condition~(\ref{eq:Sakinf}).   In~\citet{FlatWallSakiadisPaper}, the same expansion is grouped by a different gauge such that the coefficients contain $\gamma$ and $C$; the values of these parameters are obtained by forcing equation~(\ref{eq:dombal}) to satisfy constraints~(\ref{eq:SakBC1}) and~(\ref{eq:SakBC2}) via Newton iteration.  In what follows, we instead determine these parameters analytically.  In order to obtain all terms of~(\ref{eq:SakAsymp}), the variable substitutions 
\begin{equation}
    F(g(\eta))=f(\eta)/C,~g(\eta)=\gamma e^{-C\eta/2},
    \label{eq:Saktrans}
\end{equation}
are applied to the original ODE~(\ref{eq:SakODE}) to obtain the transformed ODE
\begin{subequations}
    \begin{equation}
        g^2F'''+3gF''+F'-FgF''-FF'=0,~\gamma\le g\le0
    \end{equation}
    with initial conditions (at $g=0$,i.e. $\eta=\infty$)
    \begin{equation}
        F(0)=1,~F'(0)=1,~F''(0)=1/2,
    \end{equation}
    \label{eq:SakIVP}
\end{subequations}
which are extracted from the first 3 terms of~(\ref{eq:SakAsymp}).  Note that the physical domain $0\le\eta<\infty$ becomes $\gamma\le g\le0$ under transformation~(\ref{eq:Saktrans}), where in writing the inequality we presume that $\gamma$ is negative---this will be confirmed in what follows. Since there are no unknown parameters in either the ODE or initial conditions of~(\ref{eq:SakIVP}), the power series solution of the transformed system~(\ref{eq:SakIVP}) may be obtained as 
\begin{subequations}
    \begin{equation}
        F=\sum_{n=0}^\infty A_n g^n,~|g|<R_g,
    \end{equation}
    where the pattern of the coefficients started in~(\ref{eq:SakAsymp}) is found to be~\cite{FlatWallSakiadisPaper,Corrig}
    \begin{equation}
        A_{n+1}=\frac{1}{n(n+1)^2}\sum_{j=1}^n j^2A_jA_{n-j+1},~n\ge1,~A_0=A_1=1,
        \label{eq:A}
    \end{equation}
    \label{eq:Sakseries}
\end{subequations}
and the radius of convergence (obtained from the root test) is $R_g\approx3.5$. The recursion~(\ref{eq:A}) indicates that all of the $A_n$ coefficients of ~(\ref{eq:Sakseries}) are positive, and Pringsheim's theorem~\cite{Markushevich} indicates that a singularity lies on the positive real line at $g\approx3.5$.  Since the physical domain of~(\ref{eq:SakIVP}) is $\gamma\le g\le0$, the series~(\ref{eq:Sakseries}) converges over the entire physical domain for $|\gamma|<R_g$; this is verified in what follows. 

\begin{figure}
    \centering
    \includegraphics[width=2.5in]{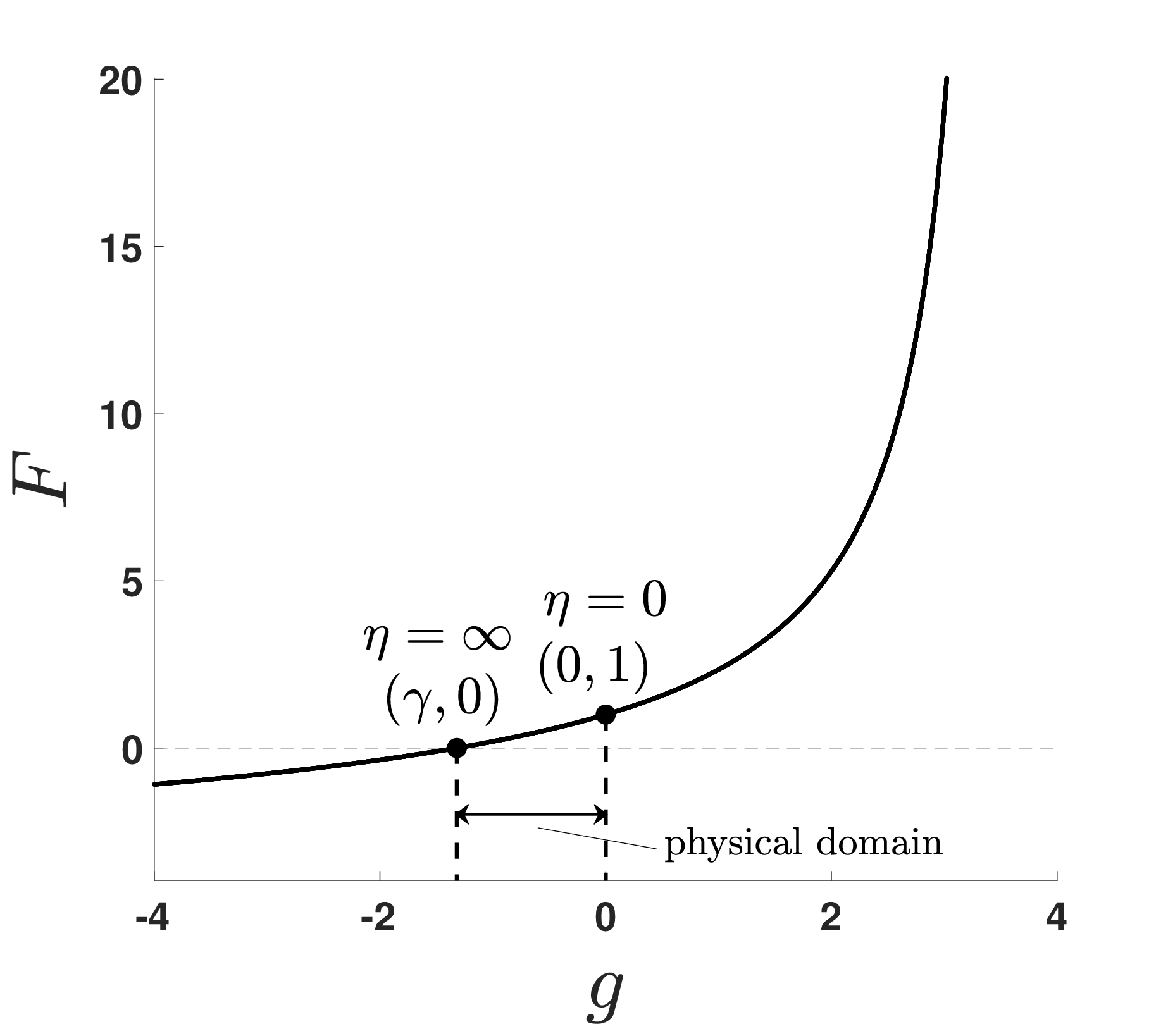}
    \caption{The numerical solution of~(\ref{eq:SakIVP}), obtained using the $4^\mathrm{th}$-order accurate Runge-Kutta scheme using a step size of $\Delta g=10^{-4}$ and taking special care to pass through removable singularities.}
    \label{fig:Fvsg}
\end{figure}

The 3 unknowns $\gamma$, $C$, and $\kappa$, may be obtained by solving the 3 equations resulting from applying the transformation~(\ref{eq:Saktrans}) to the (so far unused) conditions~(\ref{eq:SakBC1}),~(\ref{eq:SakBC2}), and~(\ref{eq:SakBC3}), leading to
\begin{subequations}
    \begin{equation}
        f(\eta=0)=CF(\gamma)=0,
    \end{equation}
    \begin{equation}
        f'(\eta=0)=-C^2\gamma F'(\gamma)/2=1,
        \label{eq:SakTransBC1}
    \end{equation}
    \begin{equation}
        f''(\eta=0)=\frac{C^3\gamma}{4}\left[F'(\gamma)+\gamma F''(\gamma)\right]=\kappa,
        \label{eq:SakTransBC2}
    \end{equation}
\end{subequations}
which, upon substituting~(\ref{eq:SakTransBC1}) into~(\ref{eq:SakTransBC2}) and simplifying, yields
\begin{subequations}
    \begin{equation}
        F(\gamma)=0,
        \label{eq:gamma}
    \end{equation}
    \begin{equation}
        C=\left[\frac{-\gamma}{2}F'(\gamma)\right]^{-1/2}
        \label{eq:SakTransBC1b}
    \end{equation}
    \begin{equation}
  \kappa=\frac{-C}{2}+\frac{C^3}{4}\gamma^2F''(\gamma).
        \label{eq:SakTransBC2b}
    \end{equation}
\end{subequations}
In arriving at~(\ref{eq:SakTransBC1b}), the positive sign of the square root has been chosen to be consistent with~(\ref{eq:C}). We begin by solving~(\ref{eq:gamma}) for $\gamma$. As seen in Fig.~\ref{fig:Fvsg}, a numerical solution of~(\ref{eq:SakIVP}) indicates that there is only one intersection of the curve $F(g)$ across the $g$-axis; thus there is only one real root $\gamma$ of~(\ref{eq:gamma}), which we now solve for via series reversion. 

A formula for series reversion developed in the Appendix is applied to~(\ref{eq:Sakseries}), and an expression for the inverse function $g(F)$ is thus obtained as: 
\begin{subequations}
\begin{equation}
g=\sum_{n=1}^\infty b_n (F-1)^n,~|F|<R_F
\label{eq:gseries}
\end{equation}
\begin{equation}
    b_1=1,~b_{n+1}=\frac{\mathbf{D}_{n,n+1}}{n+1},
    \label{eq:b}
    \end{equation}
    where the $\mathbf{D}_{n,m}$ coefficients are recursively computed as 
    \begin{eqnarray}
    \nonumber
        \mathbf{D}_{1,m}&=&-mA_2,\\ \nonumber
        \mathbf{D}_{2,m}&=&-mA_3-\frac{1}{2}\left(m+1\right)A_2\mathbf{D}_{1,m},\\ \nonumber
        \mathbf{D}_{3,m}&=&-mA_4-\frac{1}{3}\left[\left(m+2\right)A_2,\mathbf{D}_{2,m}+\left(2m+1\right)A_3\mathbf{D}_{1,m}\right],\\ \nonumber
        \vdots \\  \nonumber
    \mathbf{D}_{n,m}&=&-mA_{n+1}-\frac{1}{n}\sum_{j=1}^{n-1}\left(jm+n-j\right)A_{j+1}\mathbf{D}_{n-j,m},\\
    m&=&1,2,3,\dots, 
    \label{eq:matrix}
\end{eqnarray}
\label{eq:revert}
\end{subequations}
and the pattern is written in~(\ref{eq:matrix}) to alert the reader that, in order to obtain $\mathbf{D}_{n,n+1}$ in~(\ref{eq:b}), previously determined terms are required\footnote{A naive substitution of $m=n+1$ directly into the expression for $\mathbf{D}_{n,m}$ in~(\ref{eq:matrix}) (for successively increased $n$) to obtain $\mathbf{D}_{1,2}$, $\mathbf{D}_{2,3}$, etc, will fail since (for example) $\mathbf{D}_{2,3}$ requires knowledge of $\mathbf{D}_{1,3}$.}.   The radius of convergence (obtained from the root test) of~(\ref{eq:gseries}) is $R_F\approx4.7$.  Applying~(\ref{eq:gamma}) to~(\ref{eq:gseries}) leads to an exact expression for $\gamma$:
\begin{eqnarray}
        \gamma&=&\sum_{n=1}^\infty b_n (-1)^n\\ \nonumber
        &=&-1.3187998049666623818339568382086\dots
        \label{eq:gamval}
        \label{eq:exactgamma}
\end{eqnarray}
The above quadruple precision value is obtained using 45 terms of the series; the double precision value is obtained using 22 terms. This value of $\gamma$ aligns with intersection in Fig.~\ref{fig:Fvsg} and assures that the series solution for $F(g)$ given by~(\ref{eq:Sakseries}) and the series solution for the inverse function $g(F)$ given by~(\ref{eq:revert}) converge over the full physical domain; that is, the radii of convergence for both series lie well beyond the endpoints of their respective physical domains.

To close the problem and convert back to $f(\eta)$ via~(\ref{eq:Saktrans}), we require a value for $C$, which we may now obtain from~(\ref{eq:SakTransBC1b}). Although we we could use~(\ref{eq:Sakseries}) to obtain $F'(\gamma)$ needed for~(\ref{eq:SakTransBC1b}), less terms are required to achieve the same accuracy in the series for the inverse function~(\ref{eq:revert}).  We employ the identity $F'(g)=1/g'(F)$ such that $F'(\gamma)=1/g'(0)$ and thus from~(\ref{eq:revert}) we have
\begin{equation}
    F'(\gamma)=\left[\sum_{n=1}^\infty n b_n (-1)^{n-1}\right]^{-1}.
    \label{eq:Fprime}
\end{equation}
Substituting~(\ref{eq:Fprime}) into~(\ref{eq:SakTransBC1b}) leads to an exact expression for $C$:
\begin{eqnarray}
    C&=&\left[\frac{-2}{\gamma}\sum_{n=1}^\infty n b_n (-1)^{n-1}\right]^{1/2}\\ \nonumber
    &=&1.6161254468046037170271174250288\dots
    \label{eq:exactC}
\end{eqnarray}
The above quadruple precision value is obtained using 50 terms of the series; the double precision value is obtained using 23 terms. 

Finally, to compute the wall shear parameter $\kappa$, we may employ the identity $F''(g)=-g''(F)/(g'(F))^3$ such that $F''(\gamma)=-g''(0)/(g'(0))^3$.  This may be simplified (again using $F'(\gamma)=1/g'(0)$ and~(\ref{eq:SakTransBC1b})) such that $F''(\gamma)=-8 g''(0)/(\gamma^3 C^6)$ and thus from~(\ref{eq:revert}) we have
\begin{equation}
    F''(\gamma)=\frac{8}{\gamma^3 C^6}\sum_{n=2}^\infty n(n-1) b_n (-1)^n.
\label{eq:ddF}
\end{equation}
Substituting~(\ref{eq:ddF}) into~(\ref{eq:SakTransBC2b}) leads to an exact expression for $\kappa$:
\begin{eqnarray}
    \kappa&=&\frac{-C}{2}+\frac{2}{\gamma C^3}\sum_{n=2}^\infty n(n-1) b_n (-1)^n\\ \nonumber
    &=&-0.4437483133688610511198328438501\dots
    \label{eq:exactkappa}
\end{eqnarray}
The above quadruple precision value is obtained using 50 terms of the series; the double precision value is obtained using 23 terms. 

To summarize, the exact analytical solution to the Sakiadis problem~(\ref{eq:SakBVP}) is given by~(\ref{eq:Sakseries}), written in terms of the original variables as
\begin{eqnarray}
\nonumber
    f&=&C\sum_{n=0}^\infty A_n \left(\gamma e^{-C\eta/2}\right)^n,~0\le\eta<\infty,\\ \nonumber
        A_{n+1}&=&\frac{1}{n(n+1)^2}\sum_{j=1}^n j^2A_jA_{n-j+1},~n\ge1,~A_0=A_1=1,\\ \nonumber
        C&=&\left[\frac{-2}{\gamma}\sum_{n=1}^\infty n b_n (-1)^{n-1}\right]^{1/2},~ \gamma=\sum_{n=1}^\infty b_n (-1)^n,\\ \nonumber
        b_1&=&1,~b_{n+1}=\frac{\mathbf{D}_{n,n+1}}{n+1},~\mathbf{D}_{1,m}=\frac{-m}{4},~m=1,2,3,\dots\\
        \mathbf{D}_{n,m}&=&-mA_{n+1}-\frac{1}{n}\sum_{j=1}^{n-1}\left(jm+n-j\right)A_{j+1}\mathbf{D}_{n-j,m}.
        \label{eq:full}
    \end{eqnarray}
As indicated in~(\ref{eq:full}) and shown explicitly in~(\ref{eq:matrix}), the evaluation of the coefficient $\mathbf{D}_{n,n+1}$ in $b_n$ requires that previous terms are first determined.  For convenience, a function that implements~(\ref{eq:full}) is posted in the MATLAB repository~\cite{code}. The solution~(\ref{eq:full}) is exact in double precision arithmetic when at least 37 terms are used in the 3 infinite series which constitute the solution. The trend of error versus series truncation may be found in~\citet{FlatWallSakiadisPaper}.



\begin{figure}
    \centering
    \includegraphics[width=4in]{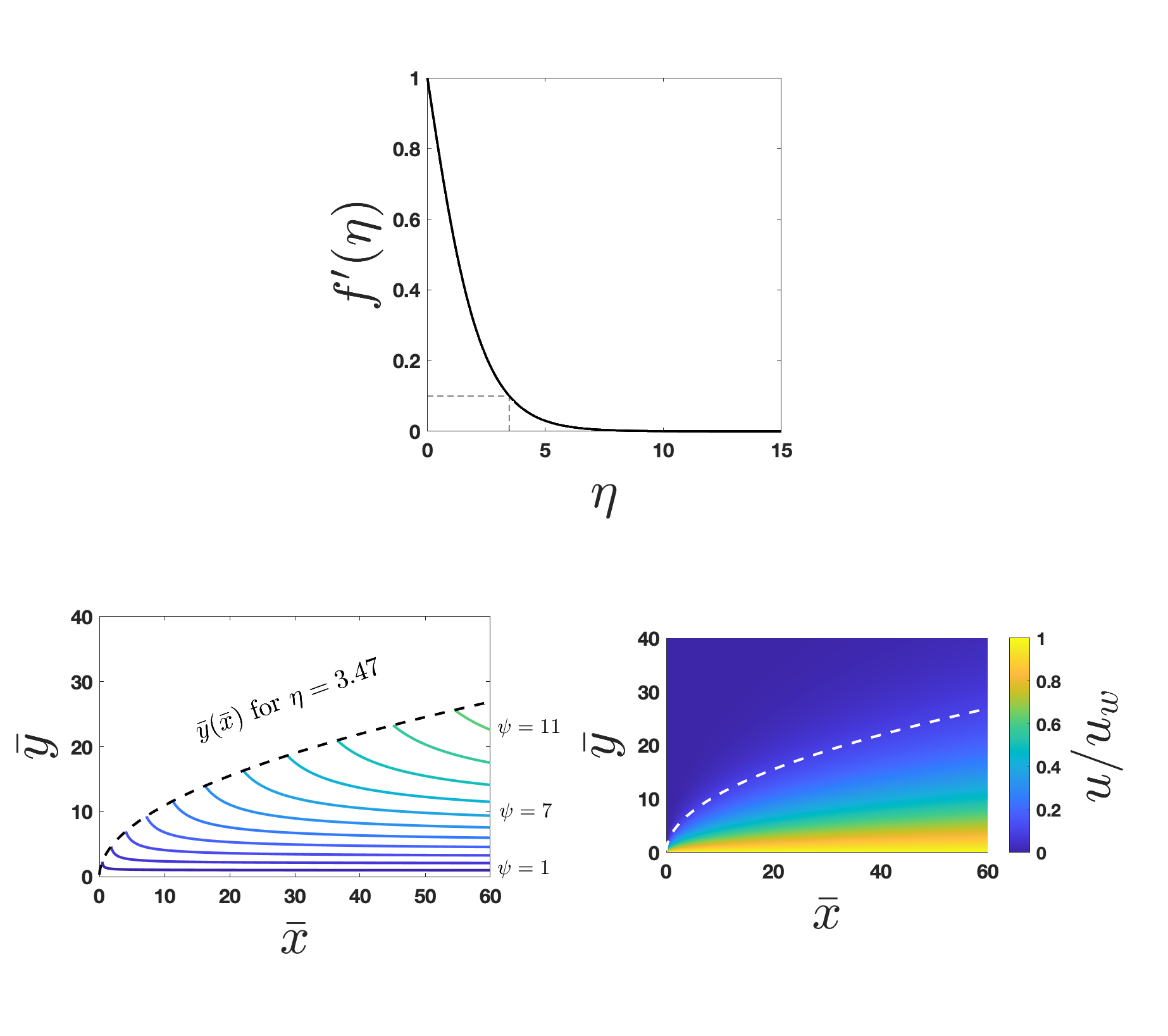}
   \caption{Example usage of the exact Sakiadis solution~(\ref{eq:full}), using 37 terms. (top) $f'(\eta)$ with gridlines indicating that $df/d\eta=u/u_w=0.1$ at $\eta\approx3.47$. (left) Contours of constant $\psi$ obtained, displayed in increments of $\Delta\psi=1$ in the $\bar{y}\equiv u_w y$ vs. $\bar{x}\equiv \nu u_w x$ plane. The dashed curve is the envelope of the boundary layer (chosen here to be the locus of points at which $u/u_w = 0.1$), and restricts the display of the streamlines for velocities where $u/u_w > 0.1$. (right) The $u$ velocity field.}
    \label{fig:SL}
\end{figure}

The streamlines of constant $\psi$ and the velocity in the x-direction (see Fig.~\ref{fig:SL}) may be extracted easily from the analytical solution by converting back to physical coordinates~\cite{Sakiadis} via 
\begin{equation}
y=\frac{\psi\eta}{u_w f(\eta)},~~x=\frac{1}{\nu u_w}\left(\frac{\psi}{f(\eta)}\right)^{2},~u=u_wf'(\eta),
\label{eq:xandy}
\end{equation}
where $u_w$ is the velocity of the moving wall and $\nu$ is kinematic viscosity; see Fig.~\ref{fig:SL}.  Note that interpreting~(\ref{eq:xandy}), $\eta$ is treated as a parameter that enables the full physical solution to be generated.   The benefit of the analytical solution is clearly indicated here, as streamline plots, can be generated accurately to any desired resolution with low computational cost.



In summary, we have determined exact values of $C$, $\gamma$, and $\kappa$, such that the analytical solution provided in~\citet{FlatWallSakiadisPaper} is now explicit; there is no longer a need for an iterative technique to find these parameters.  It is natural to ask if the same can be done for the (arguably more famous) Blasius boundary layer problem for flow over a stationary plate~\cite{Blasius}.  Differences in the $\eta\to\infty$ behaviors of the Blasius and Sakiadis problems are outlined in~\citet{FlatWallSakiadisPaper}, including the reasons why the Blasius expansion analagous to~(\ref{eq:dombal}) is not as straightforward to utilize. The Eulerized expansion\footnote{Eulerization~\cite{VanDyke} is a specific resummation technique, for functions with specific singularity orientations. that effectively transforms a divergent series into a convergent series by mapping the influence of the singularity to outside of the transformed physical domain.  Although Eulerization, when applied to the original divergent Blasius series, amplifies round-off error, this may be circumvented by instead writing the Blasius ODE and boundary conditions in terms of the Eulerized variable from the outset~\cite{SeriesBook}; this has been recently established for the meniscus on the outside of a cylindrical wall in~\cite{Cylinder}.} given by~\citet{Boyd1999} is an exact analytical solution to the Blasius problem, once an exact analytical expression (akin to~(\ref{eq:exactkappa})) for its $\kappa$ value (which is an input parameter to Boyd's solution) is obtained.


\appendix
\section*{Appendix: Recursive Formulae for Series Reversion}
The implementation below follows the developments of~\citet{Henrici}. Given a power series
\begin{equation}
    y=\sum_{n=0}^\infty a_n x^n, a_1\neq0,
\end{equation}
and the expansion of the inverse function given by
\begin{equation}
    x=\sum_{n=1}^\infty b_n \left(y-a_0\right)^n,
\end{equation}
the process of \textit{series reversion} refers to the determination of the $b_n$ coefficients, given the $a_n$ coefficients. This process may be done recursively via the combination of Lagrange's expansion formula (Equation 3.6.6 in~\citet{Abramowitz}),
\begin{equation}
    b_{n+1}=\frac{1}{\left(n+1\right)!}\left[\frac{d^{n}\left(\tilde{y}^{-n-1}\right)}{dx^{n}}\right]_{x=0},~\tilde{y}=\frac{y-a_0}{x},
    \label{eq:Taylor}
\end{equation}
with the formula (due to~\citet{Euler} and~\citet{Hansted}, a.k.a. the J.C.P. Miller Formula~\cite{Henrici}) for raising a series to a power,
\begin{subequations}
\begin{equation}
\tilde{y}^{-m}=\left(\sum_{n=0}^\infty a_{n+1} x^n\right)^{-m}=\sum_{n=0}^\infty \mathbf{D}_{n,m} x^n,
\label{eq:JCP}
\end{equation}
\begin{equation}
\mathbf{D}_{n>0,m}=\frac{-1}{na_1}\sum_{j=1}^n\left(jm+n-j\right)a_{j+1}\mathbf{D}_{n-j,m},
\label{eq:nonzero}
\end{equation}
\begin{equation}
    \mathbf{D}_{0,m}=a_1^{-m}.
    \label{eq:zero}
\end{equation}
\end{subequations}
The Taylor coefficients of~(\ref{eq:JCP}) for $m=n+1$ are exactly given by~(\ref{eq:Taylor}) multiplied by $(n+1)$, such that
\begin{equation}
    b_{n+1}=\frac{\textbf{D}_{n,n+1}}{n+1}.
    \label{eq:frac}
\end{equation}
Substitution of~(\ref{eq:zero}) into~(\ref{eq:frac}) for $n=0$ yields 
\begin{subequations}
\begin{equation}
    b_1=\frac{1}{a_1},
    \label{eq:b1}
\end{equation}
and substitution of~(\ref{eq:nonzero}) into~(\ref{eq:frac}) for $n>0$ yields
\begin{eqnarray}
    \nonumber 
        b_{n+1}&=&\frac{\textbf{D}_{n,n+1}}{n+1},~n>0\\
        \nonumber
\mathbf{D}_{n,m}&=&\frac{-ma_{n+1}}{a_1^{m+1}}-\frac{1}{na_1}\sum_{j=1}^{n-1}\left(jm+n-j\right)a_{j+1}\textbf{D}_{n-j,m},\\
m&=&1,2,3,\dots,
\label{eq:bn}
\end{eqnarray}
\end{subequations}
where the 2$^\mathrm{nd}$ term in~(\ref{eq:bn}) is interpreted as 0 for $n=1$.  Equations~(\ref{eq:b1}) and~(\ref{eq:bn}) constitute the general reversion formulae and are used in the main text to arrive at~(\ref{eq:revert}).

\bibliography{SakiadisArxiv}

\providecommand{\noopsort}[1]{}\providecommand{\singleletter}[1]{#1}%
\begin{thebibliography}{22}%
\makeatletter
\providecommand \@ifxundefined [1]{%
 \@ifx{#1\undefined}
}%
\providecommand \@ifnum [1]{%
 \ifnum #1\expandafter \@firstoftwo
 \else \expandafter \@secondoftwo
 \fi
}%
\providecommand \@ifx [1]{%
 \ifx #1\expandafter \@firstoftwo
 \else \expandafter \@secondoftwo
 \fi
}%
\providecommand \natexlab [1]{#1}%
\providecommand \enquote  [1]{``#1''}%
\providecommand \bibnamefont  [1]{#1}%
\providecommand \bibfnamefont [1]{#1}%
\providecommand \citenamefont [1]{#1}%
\providecommand \href@noop [0]{\@secondoftwo}%
\providecommand \href [0]{\begingroup \@sanitize@url \@href}%
\providecommand \@href[1]{\@@startlink{#1}\@@href}%
\providecommand \@@href[1]{\endgroup#1\@@endlink}%
\providecommand \@sanitize@url [0]{\catcode `\\12\catcode `\$12\catcode
  `\&12\catcode `\#12\catcode `\^12\catcode `\_12\catcode `\%12\relax}%
\providecommand \@@startlink[1]{}%
\providecommand \@@endlink[0]{}%
\providecommand \url  [0]{\begingroup\@sanitize@url \@url }%
\providecommand \@url [1]{\endgroup\@href {#1}{\urlprefix }}%
\providecommand \urlprefix  [0]{URL }%
\providecommand \Eprint [0]{\href }%
\providecommand \doibase [0]{https://doi.org/}%
\providecommand \selectlanguage [0]{\@gobble}%
\providecommand \bibinfo  [0]{\@secondoftwo}%
\providecommand \bibfield  [0]{\@secondoftwo}%
\providecommand \translation [1]{[#1]}%
\providecommand \BibitemOpen [0]{}%
\providecommand \bibitemStop [0]{}%
\providecommand \bibitemNoStop [0]{.\EOS\space}%
\providecommand \EOS [0]{\spacefactor3000\relax}%
\providecommand \BibitemShut  [1]{\csname bibitem#1\endcsname}%
\let\auto@bib@innerbib\@empty
\bibitem [{\citenamefont {Weinstein}\ and\ \citenamefont
  {Ruschak}(2004)}]{Weinstein2004}%
  \BibitemOpen
  \bibfield  {author} {\bibinfo {author} {\bibfnamefont {S.~J.}\ \bibnamefont
  {Weinstein}}\ and\ \bibinfo {author} {\bibfnamefont {K.~J.}\ \bibnamefont
  {Ruschak}},\ }\bibfield  {title} {\enquote {\bibinfo {title} {Coating
  flows},}\ }\href@noop {} {\bibfield  {journal} {\bibinfo  {journal} {Ann.
  Rev. Fluid Mech.}\ }\textbf {\bibinfo {volume} {36}},\ \bibinfo {pages}
  {29--53} (\bibinfo {year} {2004})}\BibitemShut {NoStop}%
\bibitem [{\citenamefont {Sakiadis}(1961)}]{Sakiadis}%
  \BibitemOpen
  \bibfield  {author} {\bibinfo {author} {\bibfnamefont {B.~C.}\ \bibnamefont
  {Sakiadis}},\ }\bibfield  {title} {\enquote {\bibinfo {title} {Boundary-layer
  behavior on continuous solid surfaces: {II} the boundary layer on a
  continuous flat surface},}\ }\href@noop {} {\bibfield  {journal} {\bibinfo
  {journal} {AlChE J.}\ }\textbf {\bibinfo {volume} {7}},\ \bibinfo {pages}
  {221--225} (\bibinfo {year} {1961})}\BibitemShut {NoStop}%
\bibitem [{\citenamefont {Hattori}(2023)}]{Hattori}%
  \BibitemOpen
  \bibfield  {author} {\bibinfo {author} {\bibfnamefont {Y.}~\bibnamefont
  {Hattori}},\ }\bibfield  {title} {\enquote {\bibinfo {title} {Numerical
  simulations of {S}akiadis boundary-layer flow},}\ }\href@noop {} {\bibfield
  {journal} {\bibinfo  {journal} {Phys. Fluids}\ }\textbf {\bibinfo {volume}
  {35}},\ \bibinfo {pages} {1--10} (\bibinfo {year} {2023})}\BibitemShut
  {NoStop}%
\bibitem [{\citenamefont {Barlow}\ \emph {et~al.}(2017)\citenamefont {Barlow},
  \citenamefont {Stanton}, \citenamefont {Hill}, \citenamefont {Weinstein},\
  and\ \citenamefont {Cio}}]{Barlow:2017}%
  \BibitemOpen
  \bibfield  {author} {\bibinfo {author} {\bibfnamefont {N.~S.}\ \bibnamefont
  {Barlow}}, \bibinfo {author} {\bibfnamefont {C.~R.}\ \bibnamefont {Stanton}},
  \bibinfo {author} {\bibfnamefont {N.}~\bibnamefont {Hill}}, \bibinfo {author}
  {\bibfnamefont {S.~J.}\ \bibnamefont {Weinstein}},\ and\ \bibinfo {author}
  {\bibfnamefont {A.~G.}\ \bibnamefont {Cio}},\ }\bibfield  {title} {\enquote
  {\bibinfo {title} {On the summation of divergent, truncated, and
  underspecified power series via asymptotic approximants},}\ }\href
  {https://doi.org/https://doi.org/10.1093/qjmam/hbw014} {\bibfield  {journal}
  {\bibinfo  {journal} {Q. J. Mech. Appl. Math.}\ }\textbf {\bibinfo {volume}
  {70}},\ \bibinfo {pages} {21--48} (\bibinfo {year} {2017})}\BibitemShut
  {NoStop}%
\bibitem [{\citenamefont {Naghshineh}\ \emph
  {et~al.}(2023{\natexlab{a}})\citenamefont {Naghshineh}, \citenamefont
  {Reinberger}, \citenamefont {Barlow}, \citenamefont {Samaha},\ and\
  \citenamefont {Weinstein}}]{FlatWallSakiadisPaper}%
  \BibitemOpen
  \bibfield  {author} {\bibinfo {author} {\bibfnamefont {N.}~\bibnamefont
  {Naghshineh}}, \bibinfo {author} {\bibfnamefont {W.~C.}\ \bibnamefont
  {Reinberger}}, \bibinfo {author} {\bibfnamefont {N.~S.}\ \bibnamefont
  {Barlow}}, \bibinfo {author} {\bibfnamefont {M.~A.}\ \bibnamefont {Samaha}},\
  and\ \bibinfo {author} {\bibfnamefont {S.~J.}\ \bibnamefont {Weinstein}},\
  }\bibfield  {title} {\enquote {\bibinfo {title} {{On the use of
  asymptotically motivated gauge functions to obtain convergent series
  solutions to nonlinear ODEs}},}\ }\href@noop {} {\bibfield  {journal}
  {\bibinfo  {journal} {IMA Journal of Applied Mathematics}\ }\textbf {\bibinfo
  {volume} {88}},\ \bibinfo {pages} {43--66} (\bibinfo {year}
  {2023}{\natexlab{a}})}\BibitemShut {NoStop}%
\bibitem [{Note1()}]{Note1}%
  \BibitemOpen
  \bibinfo {note} {The same approach was used to solve the analogous
  non-Newtonian problem~\cite {NN}.}\BibitemShut {Stop}%
\bibitem [{\citenamefont {Bataller}(2010)}]{Cortell}%
  \BibitemOpen
  \bibfield  {author} {\bibinfo {author} {\bibfnamefont {C.~R.}\ \bibnamefont
  {Bataller}},\ }\bibfield  {title} {\enquote {\bibinfo {title} {Numerical
  comparisons of {B}lasius and {S}akiadis flows},}\ }\href
  {https://doi.org/10.11113/MATEMATIKA.V26.N.562} {\bibfield  {journal}
  {\bibinfo  {journal} {MATEMATIKA}\ }\textbf {\bibinfo {volume} {26}},\
  \bibinfo {pages} {187--196} (\bibinfo {year} {2010})}\BibitemShut {NoStop}%
\bibitem [{\citenamefont {Naghshineh}\ \emph
  {et~al.}(2023{\natexlab{b}})\citenamefont {Naghshineh}, \citenamefont
  {Reinberger}, \citenamefont {Barlow}, \citenamefont {Samaha},\ and\
  \citenamefont {Weinstein}}]{Corrig}%
  \BibitemOpen
  \bibfield  {author} {\bibinfo {author} {\bibfnamefont {N.}~\bibnamefont
  {Naghshineh}}, \bibinfo {author} {\bibfnamefont {W.~C.}\ \bibnamefont
  {Reinberger}}, \bibinfo {author} {\bibfnamefont {N.~S.}\ \bibnamefont
  {Barlow}}, \bibinfo {author} {\bibfnamefont {M.~A.}\ \bibnamefont {Samaha}},\
  and\ \bibinfo {author} {\bibfnamefont {S.~J.}\ \bibnamefont {Weinstein}},\
  }\bibfield  {title} {\enquote {\bibinfo {title} {{Correction to: On the use
  of asymptotically motivated gauge functions to obtain convergent series
  solutions to nonlinear ODEs}},}\ }\href@noop {} {\bibfield  {journal}
  {\bibinfo  {journal} {IMA Journal of Applied Mathematics}\ }\textbf {\bibinfo
  {volume} {88}},\ \bibinfo {pages} {644} (\bibinfo {year}
  {2023}{\natexlab{b}})}\BibitemShut {NoStop}%
\bibitem [{\citenamefont {Markushevich}(1985)}]{Markushevich}%
  \BibitemOpen
  \bibfield  {author} {\bibinfo {author} {\bibfnamefont {A.~I.}\ \bibnamefont
  {Markushevich}},\ }\enquote {\bibinfo {title} {Theory of functions of a
  complex variable (three volumes in one): 2$^\mathrm{nd}$ edition},}\ \
  (\bibinfo  {publisher} {Chelsea},\ \bibinfo {year} {1985})\BibitemShut
  {NoStop}%
\bibitem [{Note2()}]{Note2}%
  \BibitemOpen
  \bibinfo {note} {A naive substitution of $m=n+1$ directly into the expression
  for $\protect \mathbf {D}_{n,m}$ in~(\ref {eq:matrix}) (for successively
  increased $n$) to obtain $\protect \mathbf {D}_{1,2}$, $\protect \mathbf
  {D}_{2,3}$, etc, will fail since (for example) $\protect \mathbf {D}_{2,3}$
  requires knowledge of $\protect \mathbf {D}_{1,3}$.}\BibitemShut {Stop}%
\bibitem [{cod()}]{code}%
  \BibitemOpen
  \href@noop {} {}\bibinfo {howpublished}
  {https://www.mathworks.com/matlabcentral/fileexchange/157956-sakiadis-function-exact-analytical-solution}\BibitemShut
  {NoStop}%
\bibitem [{\citenamefont {Blasius}(1908)}]{Blasius}%
  \BibitemOpen
  \bibfield  {author} {\bibinfo {author} {\bibfnamefont {H.}~\bibnamefont
  {Blasius}},\ }\bibfield  {title} {\enquote {\bibinfo {title} {Grenzschichten
  in flussigkeiten mit kleiner reibung},}\ }\href@noop {} {\bibfield  {journal}
  {\bibinfo  {journal} {Zeitschrift fur Mathematik und Physik}\ }\textbf
  {\bibinfo {volume} {56}},\ \bibinfo {pages} {1--37} (\bibinfo {year}
  {1908})}\BibitemShut {NoStop}%
\bibitem [{Note3()}]{Note3}%
  \BibitemOpen
  \bibinfo {note} {Eulerization~\cite {VanDyke} is a specific resummation
  technique, for functions with specific singularity orientations. that
  effectively transforms a divergent series into a convergent series by mapping
  the influence of the singularity to outside of the transformed physical
  domain. Although Eulerization, when applied to the original divergent Blasius
  series, amplifies round-off error, this may be circumvented by instead
  writing the Blasius ODE and boundary conditions in terms of the Eulerized
  variable from the outset~\cite {SeriesBook}; this has been recently
  established for the meniscus on the outside of a cylindrical wall in~\cite
  {Cylinder}.}\BibitemShut {Stop}%
\bibitem [{\citenamefont {Boyd}(1999)}]{Boyd1999}%
  \BibitemOpen
  \bibfield  {author} {\bibinfo {author} {\bibfnamefont {J.~P.}\ \bibnamefont
  {Boyd}},\ }\bibfield  {title} {\enquote {\bibinfo {title} {The {B}lasius
  function in the complex plane},}\ }\href@noop {} {\bibfield  {journal}
  {\bibinfo  {journal} {Exper. Math.}\ }\textbf {\bibinfo {volume} {8}},\
  \bibinfo {pages} {381--394} (\bibinfo {year} {1999})}\BibitemShut {NoStop}%
\bibitem [{\citenamefont {Henrici}(1956)}]{Henrici}%
  \BibitemOpen
  \bibfield  {author} {\bibinfo {author} {\bibfnamefont {P.}~\bibnamefont
  {Henrici}},\ }\bibfield  {title} {\enquote {\bibinfo {title} {Automatic
  computations with power series},}\ }\href
  {https://doi.org/https://doi.org/10.1145/320815.320819} {\bibfield  {journal}
  {\bibinfo  {journal} {JACM}\ }\textbf {\bibinfo {volume} {3}},\ \bibinfo
  {pages} {10--15} (\bibinfo {year} {1956})}\BibitemShut {NoStop}%
\bibitem [{\citenamefont {Abramowitz}\ and\ \citenamefont
  {Stegun}(1972)}]{Abramowitz}%
  \BibitemOpen
  \bibfield  {author} {\bibinfo {author} {\bibfnamefont {M.}~\bibnamefont
  {Abramowitz}}\ and\ \bibinfo {author} {\bibfnamefont {I.~A.}\ \bibnamefont
  {Stegun}},\ }\enquote {\bibinfo {title} {Handbook of mathematical
  functions},}\ \ (\bibinfo  {publisher} {Dover},\ \bibinfo {year}
  {1972})\BibitemShut {NoStop}%
\bibitem [{\citenamefont {Euler}(1748)}]{Euler}%
  \BibitemOpen
  \bibfield  {author} {\bibinfo {author} {\bibfnamefont {L.}~\bibnamefont
  {Euler}},\ }\enquote {\bibinfo {title} {Introductio in analysin
  infinitorum},}\ \ (\bibinfo  {publisher} {Lausanne},\ \bibinfo {year}
  {1748})\BibitemShut {NoStop}%
\bibitem [{\citenamefont {Hansted}(1881)}]{Hansted}%
  \BibitemOpen
  \bibfield  {author} {\bibinfo {author} {\bibfnamefont {B.}~\bibnamefont
  {Hansted}},\ }\bibfield  {title} {\enquote {\bibinfo {title} {Nogle
  bemaerkninger om bestemmelsen af koefficienterne i m'te potens af en
  potensraekke},}\ }\href@noop {} {\bibfield  {journal} {\bibinfo  {journal}
  {Tidskrift for Matematik}\ }\textbf {\bibinfo {volume} {5}},\ \bibinfo
  {pages} {236--237} (\bibinfo {year} {1881})}\BibitemShut {NoStop}%
\bibitem [{\citenamefont {Naghshineh}\ \emph
  {et~al.}(2023{\natexlab{c}})\citenamefont {Naghshineh}, \citenamefont
  {Barlow}, \citenamefont {Samaha},\ and\ \citenamefont {Weinstein}}]{NN}%
  \BibitemOpen
  \bibfield  {author} {\bibinfo {author} {\bibfnamefont {N.}~\bibnamefont
  {Naghshineh}}, \bibinfo {author} {\bibfnamefont {N.~S.}\ \bibnamefont
  {Barlow}}, \bibinfo {author} {\bibfnamefont {M.~A.}\ \bibnamefont {Samaha}},\
  and\ \bibinfo {author} {\bibfnamefont {S.~J.}\ \bibnamefont {Weinstein}},\
  }\bibfield  {title} {\enquote {\bibinfo {title} {{Asymptotically-consistent
  analytical solutions for the non-Newtonian Sakiadis boundary layer}},}\
  }\href@noop {} {\bibfield  {journal} {\bibinfo  {journal} {Physics of
  Fluids}\ }\textbf {\bibinfo {volume} {35}},\ \bibinfo {pages} {1--20}
  (\bibinfo {year} {2023}{\natexlab{c}})}\BibitemShut {NoStop}%
\bibitem [{\citenamefont {Van~Dyke}(1964)}]{VanDyke}%
  \BibitemOpen
  \bibfield  {author} {\bibinfo {author} {\bibfnamefont {M.}~\bibnamefont
  {Van~Dyke}},\ }\href@noop {} {\emph {\bibinfo {title} {Perturbation Methods
  in Fluid Mechanics}}}\ (\bibinfo  {publisher} {Academic Press, New York and
  London},\ \bibinfo {year} {1964})\BibitemShut {NoStop}%
\bibitem [{\citenamefont {Barlow}\ and\ \citenamefont
  {Weinstein}(202X)}]{SeriesBook}%
  \BibitemOpen
  \bibfield  {author} {\bibinfo {author} {\bibfnamefont {N.~S.}\ \bibnamefont
  {Barlow}}\ and\ \bibinfo {author} {\bibfnamefont {S.~J.}\ \bibnamefont
  {Weinstein}},\ }\enquote {\bibinfo {title} {Power series solutions to
  nonlinear ordinary differential equations},}\ \ (\bibinfo  {publisher} {draft
  with publisher},\ \bibinfo {year} {202X})\ Chap.\ \bibinfo {chapter} {IV:
  Improving and using power series solutions to nonlinear ODEs}\BibitemShut
  {NoStop}%
\bibitem [{\citenamefont {Naghshineh}\ \emph {et~al.}(2024)\citenamefont
  {Naghshineh}, \citenamefont {Reinberger}, \citenamefont {Barlow},
  \citenamefont {Samaha},\ and\ \citenamefont {Weinstein}}]{Cylinder}%
  \BibitemOpen
  \bibfield  {author} {\bibinfo {author} {\bibfnamefont {N.}~\bibnamefont
  {Naghshineh}}, \bibinfo {author} {\bibfnamefont {W.~C.}\ \bibnamefont
  {Reinberger}}, \bibinfo {author} {\bibfnamefont {N.~S.}\ \bibnamefont
  {Barlow}}, \bibinfo {author} {\bibfnamefont {M.~A.}\ \bibnamefont {Samaha}},\
  and\ \bibinfo {author} {\bibfnamefont {S.~J.}\ \bibnamefont {Weinstein}},\
  }\bibfield  {title} {\enquote {\bibinfo {title} {{The shape of an
  axisymmetric meniscus in a static liquid pool: effective implementation of
  the {E}uler transformation}},}\ }\href {https://doi.org/DOI:
  10.1093/imamat/hxad037} {\bibfield  {journal} {\bibinfo  {journal} {IMA
  Journal of Applied Mathematics}\ } (\bibinfo {year} {2024}),\ DOI:
  10.1093/imamat/hxad037}\BibitemShut {NoStop}%
\end{thebibliography}%

\end{document}